\documentclass[aps,prl,superscriptaddress,amsmath,amssymb,floatfix,twocolumn]{revtex4-1}
\usepackage{graphicx}
\usepackage{subfigure}
\usepackage{color}
\usepackage{epstopdf}
\usepackage{ulem}

\begin{document}

\title{Orbital-selective correlations for topology in FeSe$_{x}$Te$_{1-x}$}

\author{Zhiguang Liao}
\affiliation{Department of Physics and Beijing Key Laboratory of Opto-electronic Functional Materials \& Micro-nano Devices,
Renmin University of China, Beijing 100872, China}

\author{Rong Yu}
\email{rong.yu@ruc.edu.cn}
\affiliation{Department of Physics and Beijing Key Laboratory of Opto-electronic Functional Materials \&
Micro-nano Devices, Renmin University of China, Beijing 100872, China}
\affiliation{Key Laboratory of Quantum State Construction and Manipulation (Ministry of Education),
Renmin University of China, Beijing, 100872, China}

\author{Jian-Xin Zhu}
\email{jxzhu@lanl.gov}
\affiliation{Theoretical Division and Center for Integrated Nanotechnologies,
Los Alamos National Laboratory, Los Alamos, New Mexico 87545, USA}

\author{Qimiao Si}
\email{qmsi@rice.edu}
\affiliation{Department of Physics \& Astronomy,
Rice Center for Quantum Materials,
Rice University, Houston, Texas 77005,USA}

\begin{abstract}
Strong correlations lead to emergent excitations at low energies.
When combined with symmetry constraints, they may produce topological  electronic states near the Fermi energy.
Within this general framework, here we address
the topological features in iron-based superconductors.
We examine the effects of orbital-selective correlations
on the band inversion in the iron chalcogenide FeSe$_{x}$Te$_{1-x}$ near
its doping of optimal superconductivity,
within a multiorbital model
and using a $U(1)$
slave spin theory.
The orbital selectivity of the quasiparticle spectral weight,
along with its counterpart of the energy level renormalization,
leads to a
band inversion and Dirac node formation
pinned to the immediate vicinity of the Fermi energy.
Our work demonstrates both the naturalness and robustness
of the topological properties in FeSe$_{x}$Te$_{1-x}$,
and
uncovers a new setting in which strong correlations and space-group symmetry
cooperate in generating strongly correlated electronic topology.
\end{abstract}

\maketitle


{\it Introduction.~} Since its discovery~\cite{Kamihara_JACS_2008}, iron-based superconductors (FeSCs) have
attracted extensive research interest because of their unconventional high-temperature superconductivity
and a rich landscape of electronic orders
~\cite{Si_Hussey2023,Yi_npjQM_2017,Bascones_CRP_2016,Hirschfeld_CRP_2016,Si_NRM_2016,Dai_RMP_2015,Dagotto_RMP_2013,Wang_Sci_2011,Johnston_AP_2010}.
These properties
originate
 from electron correlations, which can be strongly orbital dependent in many FeSCs~\cite{Yu_PRB_2011,Yin_NP_2011, Yu_PRL_2013, deMedici_PRL_2014, Backes_PRB_2015, Moreo_CP_2019, Yi_PRL_2013, Wang_NC_2014, Ding_PRB_2014, Yi_NC_2015,Liu_PRB_2015, WangM_PRB_2015, Niu_Feng_PRB_2016, Hiraishi_Hosono_2020,Haihu_PRX_2020}.
The Hund's rule coupling plays a crucial role in suppressing the interorbital correlations~\cite{Yu_PRB_2011,deMedici_PRL_2014}
and drives the system towards an orbital-selective Mott phase. Strong orbital selectivity causes
a substantial renormalization of the band structure in the normal state, giving rise to a large effective mass enhancement~\cite{Huang_CP_2022,Yi_NC_2015,Yi_PRL_2013}
and reconstruction of band structure and Fermi surface~\cite{Huang_CP_2022, Lin_arXiv_2021},
and
stabilizing orbital-selective pairing states
in the superconducting state~\cite{Yu_PRB:2014,Yin_NP:2014, Ong_PNAS:2016,Nica_npjQM_2017}.
It has been recognized that strong orbital selectivity is universal in iron chalcogenides~\cite{Yi_NC_2015},
and an orbital-selective Mott transition (OSMT) can be approached
by successively increasing the Te concentration in FeSe$_{x}$Te$_{1-x}$~\cite{Huang_CP_2022}.
The orbital selectivity can be quantified by the difference in the the enhancement of the effective mass $m^{\ast}$
 from the band mass $m_b$. For FeSe$_{x}$Te$_{1-x}$ near $x=0.5$, the enhancement factor $m^{\ast}/m_b$
 is very large for $3d_{xy}$ states ($\sim$$15$) and
 moderate for 3d$_{xz/yz}$ states ($\sim$$4$), which dominate over the $e_g$ states near the Fermi energy
 ~\cite{Yi_npjQM_2017,Si_NRM_2016,Liu_PRB_2015,Yi_NC_2015,Huang_CP_2022}.

The FeSCs are being pursued as promising materials for topological superconductivity and, thus,
a candidate platform for fault-tolerant quantum computation~\cite{Nayak_RMP_2008,Kitaev_2003}.
In the iron chalcogenides
FeSe$_{x}$Te$_{1-x}$ near
$x\approx 0.5$, there has been emerging experimental evidence for
topological surface states and Majorana zero modes in vortex cores in the superconducting state
~\cite{Wang_Science_2018, Zhu_Science_2020}.
Instead of utilizing intrinsic unconventional pairing such as chiral $p$-wave pairing,
as being pursued in correlated materials such as UTe$_2$~\cite{Jiao_Nat_2020},
here one relies on the (in-situ) proximity between superconductivity and band inversion
\cite{Fu-Kane_prl2008}.
This mechanism requires that the topological band inversion takes place in the immediate vicinity of the Fermi energy
\cite{Hu-Zhang_prl22}.
The experimental evidence
is consistent with this requirement, with the surface Dirac states evinced
within about $20$ meV of the Fermi energy~\cite{Zhang_Science_2018, Zhang_NP_2019, Chen_CPL_2019, Lohani_PRB_2020, Day_PRB_2022, Li_Nat_2022}.
However, most existing theories  on the non-trivial band topology ~\cite{Wang_PRB_2015, Xu_PRL_2016, Hao_NSR_2019, Jiang_PRX_2019, Zhang_PRL_2019,Konig_prl_2019} are
anchored by
 the density functional theory (DFT) calculations,
 which found band inversion much further away from the Fermi energy, by an order of magnitude
~\cite{Wang_PRB_2015, Xu_PRL_2016}.

In this Letter, we address the above issue by showing that
the orbital-selective Mott correlations produce single-particle states that are pinned very close to $E_{\rm F}$
and
retain
band inversion and Dirac nodes along
the $\Gamma$-$Z$ ($k_z$) direction.
As such, the energies
of topological gap and Dirac node locations as measured with respect to the Fermi energy
are very small, within about $20$ meV, which
is consistent both with the experimental observation and with the requirement for surface topological superconductivity.
We reach this conclusion by investigating
a multiorbital Hubbard model including both Fe $d$ and Se/Te $p$ orbitals for FeSe$_x$Te$_{1-x}$,
using the $U(1)$ slave spin theory~\cite{Yu_PRB_2012, Yu_PRB_2017}.
The band inversion takes place between a $p_z$-$d_{xy}$ hybridized band and a $d_{xz/yz}$ band,
leading to a topological insulator-like feature in the presence of the spin-orbit coupling/
While above the gap, a Dirac node emerges which reflects the protection of a $C_{4z}$
rotational symmetry along the high symmetry line $\Gamma$-$Z$.
Still,
we show that
this picture
has a degree of robustness, because
 the orbital selective renormalization
involves
not only the quasiparticle spectral weight
but also the
energy levels.
The two types of renormalization effects
cause opposite shifts of the $d_{xy}$ and $d_{xz/yz}$ bands,
and it follows that the Fermi-energy-pinned band inversion and Dirac node formation occur
 over a wide interaction parameter range.
When the interaction becomes too large and
 the system approaches an OSMT, we find a topological phase transition to where the
 renormalized bands can no longer invert. Our results illustrate a general principle, namely
 (orbital-selective) electron correlations cooperate with space group symmetry
 to produce correlated topological properties
 \cite{Chen-Natphys22}.

  \begin{figure}[t!]
\centering\includegraphics[
width=80mm 
]{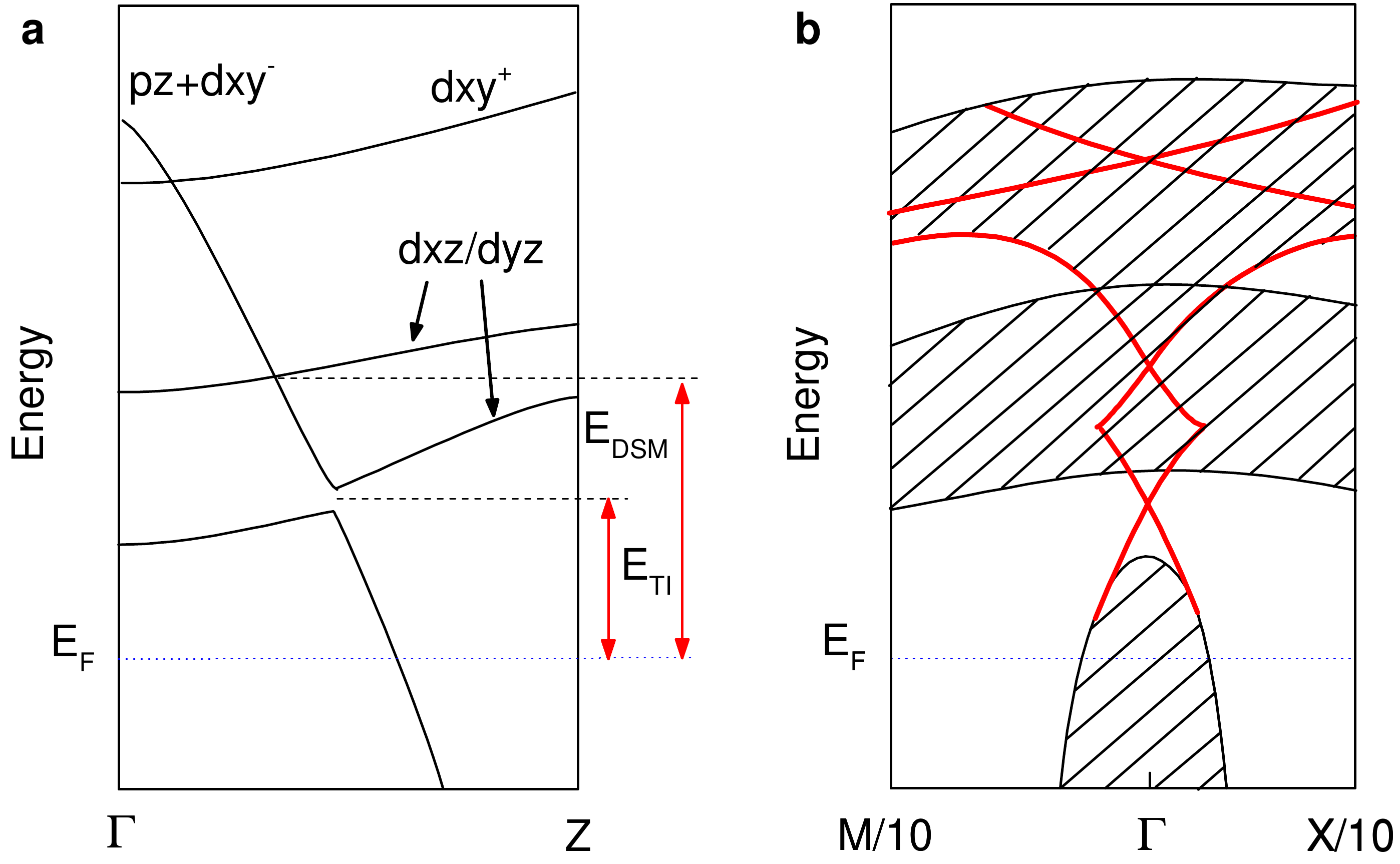}
\caption{(Color online)
{\bf a}: Sketch of the band inversion that yields a topological-insulator gap
and symmetry-protected Dirac crossing
along the $\Gamma$-$Z$ direction of the BZ in FeSe$_x$Te$_{1-x}$ at $x=0.5$,
defining two energy scales, $E_{\rm TI}$ and $E_{\rm DSM}$.
{\bf b}: Corresponding topological surface states (solid lines) near the
$\Gamma$ point. The bulk bands are
represented by
the shading.
}
\label{fig:1}
\end{figure}

{\it Qualitative picture.~}
The overall picture is illustrated in
 Fig.\,\ref{fig:1}{\bf a}.
 A topological gap caused by the band inversion is located at $E_{\rm TI}$ away from the Fermi energy ($E_{\rm F}$),
 while a
 Dirac node protected by the $C_{4z}$ lattice rotational symmetry
 is at $E_{\rm DSM}$;
 each
 gives rise to a surface Dirac-cone
 as depicted in Fig.~\ref{fig:1}{\bf b}.

\begin{figure}[t!]
\centering\includegraphics[
width=85mm 
]{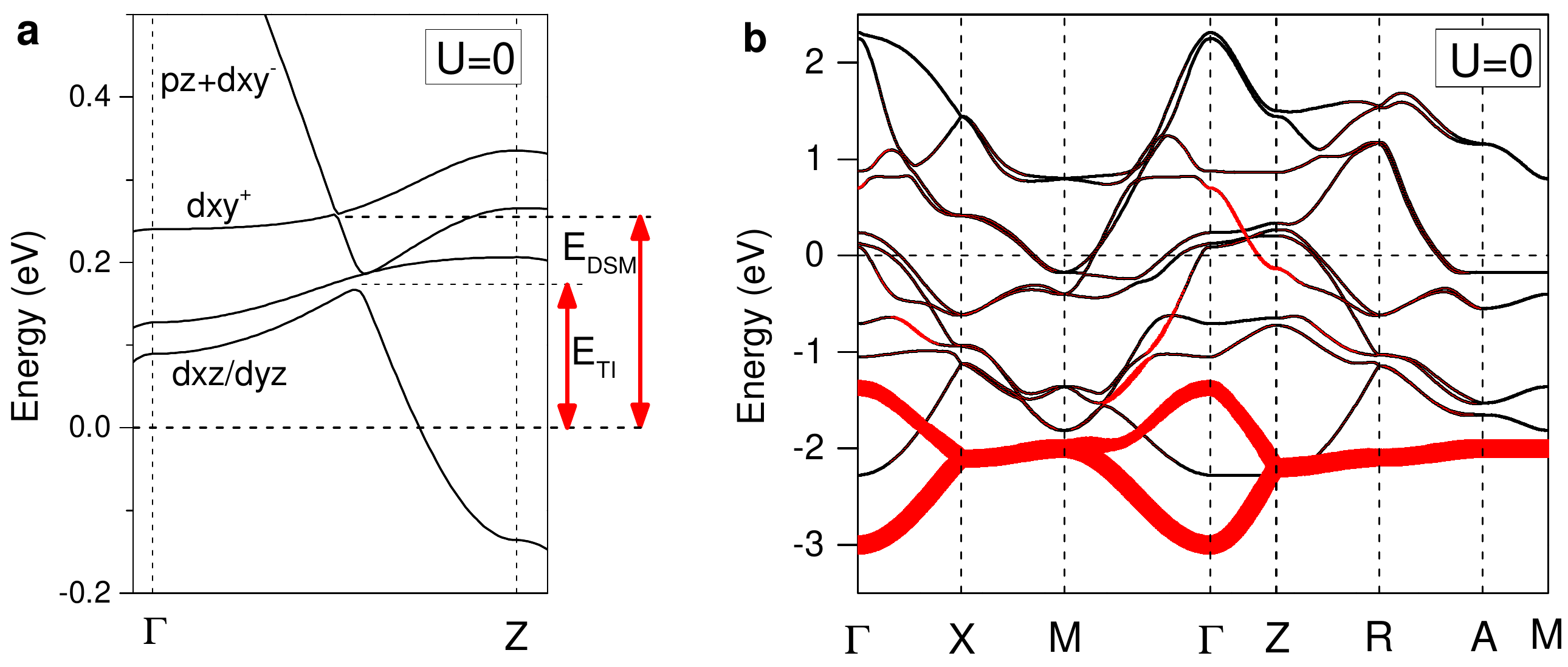}
\caption{(Color online)
Band structure of the 12-orbital Hubbard model
in the non-interacting limit with $\lambda^0_{\rm{SOC}}=60$ meV.
{\bf a}: The band structure near the Fermi energy $E_F$ (which is set to be zero) along the $\Gamma$-$Z$ direction of the BZ
showing the band inversion, which causes a bulk gap at energy position $E_{\rm{TI}}\approx170$ meV and
a
Dirac
level crossing at $E_{\rm{DSM}}\approx260$ meV.
{\bf b}: Full band structure of the 12-orbital model along high symmetric directions of the BZ.
The thickness of the red line is proportional to the $p_z$ orbital weight of the bands.
}
\label{fig:2}
\end{figure}

From the DFT calculation,
the band inversion is between a $p_z$ band and the $d_{xz/yz}$ bands along the $\Gamma$-$Z$ ($k_z$) direction of the Brillouin zone (BZ).
At the $\Gamma$ point the $p_z$ band is located at a higher energy than the $d_{xz/yz}$ bands. This is inverted at the $Z$ point
as the $p_z$ band is strongly dispersive along the $k_z$ direction.
A finite spin-orbit coupling (SOC) causes a splitting of the doubly degenerate $d_{xz/yz}$ bands, and leads to a bulk gap between the $p_z$ band
and one of the $d_{xz/yz}$ bands.
The opposite parities of the $p_z$ and $d_{xz/yz}$ bands give rise to topological surface states.
A Dirac cone-type surface state appears at the $\Gamma$ point on the (001) surface.
This is similar to the case of a topological insulator (TI) which has a bulk gap
and a Dirac cone-type gapless surface state.
Above the bulk gap, the $p_z$ band crosses the other $d_{xz/yz}$ band at a higher energy.
The level crossing is protected by the $C_{4z}$ rotational symmetry of the lattice,
and this leads to another Dirac semi-metal (DSM)-type surface state at a slightly higher energy.
In DFT calculations, the energy positions of the bulk gap and the Dirac cones are about
$200$ meV higher than the Fermi energy $E_F$
({\it c.f.} Figs.\,\ref{fig:2}{\bf a}).
The role of the electron correlations on the band topological features has been discussed~\cite{Yu_FrontPhys_21},
and recent calculations based on DFT in combination with dynamical mean-field theory (DMFT) have
suggested that
the overall correlation effects may move
such features towards the Fermi energy
~\cite{Ma_PRB_22}.

{\it Model and method.~} We study a multiorbital Hubbard model for FeSe$_x$Te$_{1-x}$,
 which includes the five $3d$ orbitals of Fe
 and the $p_z$ orbital of Se/Te. The model Hamiltonian reads as
\begin{equation}\label{HamTot}
 H = H_{\rm{TB}} + H_{\rm{soc}} + H_{\rm{int}}.
\end{equation}
Here, $H_{\rm{TB}}$ is a tight-binding Hamiltonian with tetragonal lattice symmetry.
\begin{eqnarray}
 \label{Eq:Ham_0}
 && H_{\rm{TB}} =\frac{1}{2}\sum_{ij\alpha\beta\sigma} t^{\alpha\beta}_{ij}
 d^\dagger_{i\alpha\sigma} d_{j\beta\sigma} + \sum_{i\alpha\sigma} (\epsilon_\alpha-\mu)
 d^\dagger_{i\alpha\sigma} d_{i\alpha\sigma} \nonumber \\
 && + \sum_{il\alpha\sigma} t^{p\alpha}_{li} (p^\dagger_{l\sigma} d_{i\alpha\sigma}
 + d^\dagger_{i\alpha\sigma} p_{l\sigma}) + \sum_{l\sigma} (\epsilon_p-\mu) p^\dagger_{l\sigma} p_{l\sigma},
\end{eqnarray}
where $d^\dagger_{i\alpha\sigma}$ creates an electron in orbital $\alpha$ ($\alpha=1,...,5$
denoting $xz$, $yz,$ $x^2-y^2$, $xy$, and $3z^2-r^2$ orbitals, respectively) with spin $\sigma$ at site $i$, $\epsilon_\alpha$
refers to the $d$-orbital energy level associated with the crystal field splittings,
$p^\dagger_{l\sigma}$ creates an electron in the $p_z$ orbital of Se/Te, $\epsilon_p$ denotes the $p_z$-orbital onsite energy,
and $\mu$ is the chemical potential.
The tight-binding parameters $t^{\alpha\beta}_{ij}$, $t^{p\alpha}_{li}$, $\epsilon_\alpha$, and $\epsilon_p$
can be obtained by fitting the calculated DFT band structure and projecting to a 12-orbital basis
including the Fe $3d$ orbitals and the $p_z$ orbital of Se/Te as described in Supplemental Material (SM)~\cite{SM}.
In addition, the atomic SOC term in the $d$-orbital sector is
$H_{\rm{soc}}= \frac{1}{2} \lambda^0_{\rm{soc}} \sum_{i\alpha\beta \sigma \sigma^{\prime}} \left(\mathbf{L}\cdot\boldsymbol{\tau}\right)_{\alpha \sigma, \beta \sigma^{\prime}} d^\dagger_{i\alpha \sigma} d_{i\beta \sigma^{\prime}}$,
where $\mathbf{L}$ denotes the orbital angular momentum operator and $\boldsymbol{\tau}$ refers to the Pauli matrices. A bare value of the SOC $\lambda^0_{\rm{soc}}=60$ meV is taken,
which will be renormalized by interactions.
We adjust the chemical potential so that the total electron density is $16$ per unit cell,
which includes $12$ electrons on Fe sites and 4 electrons on Se/Te sites in a unit cell containing two Fe and two Se/Te ions.
The on-site interaction $H_{\rm{int}}$ reads
\begin{eqnarray}
 \label{Eq:Ham_int} H_{\rm{int}} &=& \frac{U}{2} \sum_{i,\alpha,\sigma}n_{i\alpha\sigma}n_{i\alpha\bar{\sigma}}\nonumber\\
 &+&\sum_{i,\alpha<\beta,\sigma} \left\{ U^\prime n_{i\alpha\sigma} n_{i\beta\bar{\sigma}}\right.
 + (U^\prime-J_{\rm{H}}) n_{i\alpha\sigma} n_{i\beta\sigma}\nonumber\\
&-&\left.J_{\rm{H}}(d^\dagger_{i\alpha\sigma}d_{i\alpha\bar{\sigma}} d^\dagger_{i\beta\bar{\sigma}}d_{i\beta\sigma}
 +d^\dagger_{i\alpha\sigma}d^\dagger_{i\alpha\bar{\sigma}}
 d_{i\beta\sigma}d_{i\beta\bar{\sigma}}) \right\},
\end{eqnarray}
where $n_{i\alpha\sigma}=d^\dagger_{i\alpha\sigma} d_{i\alpha\sigma}$.
Here,
$U$, $U^\prime$, and $J_{\rm{H}}$, respectively denote the intra-
and inter- orbital repulsion and the Hund's rule coupling, and
$U^\prime=U-2J_{\rm{H}}$
is taken~\cite{Castellani_PRB_1978}. In the calculation, we fix $J_{\rm{H}}/U=0.25$ and neglect interactions in the $p$ orbitals.

The correlation effects of the above model are investigated by using a $U(1)$ slave-spin theory~\cite{Yu_PRB_2012, Yu_PRB_2017}.
In the $U(1)$ slave-spin theory, the $d$-electron operator is rewritten as
$d^\dagger_{i\alpha\sigma} = S^+_{i\alpha\sigma} f^\dagger_{i\alpha\sigma}$, where $S^+_{i\alpha\sigma}$ ($f^\dagger_{i\alpha\sigma}$) is a quantum $S=1/2$ spin (fermionic spinon) operator introduced to carry the
electron's charge (spin) degree of freedom, and $S^z_{i\alpha\sigma} = f^\dagger_{i\alpha\sigma} f_{i\alpha\sigma} - \frac{1}{2}$ is a local constraint.
At the saddle-point level, we employ a Lagrange multiplier $\lambda_{\alpha}$
to handle the constraint, and decompose the slave-spin and spinon operators. In this way, the Hamiltonian in Eqn.~\eqref{HamTot} is solved by determining $\lambda_{\alpha}$ and the quasiparticle spectral weight $Z_\alpha\propto |\langle S^+_{\alpha} \rangle|^2$ self-consistently~\cite{Yu_PRB_2012, Yu_PRB_2017}.

{\it Band topology under orbital-selective correlations.~}
We focus on studying the band structure of the above model for FeSe$_x$Te$_{1-x}$ at $x=0.5$,
where the non-trivial band topology is observed in experiments. Fig.~\ref{fig:2}{\bf b}
shows the band structure of the 12-orbital model with a finite SOC
in the non-interacting limit $U=J_{\rm{H}}=0$.
Bands across the Fermi energy have mainly the $d$-orbital character except one band along the $\Gamma$-$Z$ direction (shown in red),
in which the dominant orbital character is $p_z$.
This band
involves the $p_z$ orbital hybridizing with the $d_{xy}$ orbital and has an odd parity.
As
 shown in Fig.~\ref{fig:2}{\bf a}, this band crosses another $d_{xy}$ band with even parity ($d_{xy}^+$)
 and two $d_{xz/yz}$ hybridized bands in turn. In particular, the crossing between the odd-parity $p_z+d_{xy}^-$ band and one even-parity $d_{xz/yz}$ band identifies a band inversion.
 The latter gives rise to $Z_2$ band topology as in a topological insulator (TI):
  In the presence of the SOC, a bulk gap develops
 whereas the surface mode keeps as a gapless Dirac cone (see below). In the calculation,
 the other $d_{xz/yz}$ band split by the SOC lies in the bulk gap. Slightly above the bulk gap, the $p_z+d_{xy}^-$ band
 crosses the $d_{xy}^+$ band, leading to a Dirac semi-metal (DSM) feature.
 We can define two characteristic energies of the band topology:
 The energy position (to $E_F$) of the bulk gap $E_{\rm{TI}}\approx 170$ meV, and the energy of the DSM-like crossing is
 $E_{\rm{DSM}}\approx260$ meV.
Although the band topology is present in the non-interacting limit, the characteristic energies are too large compared to those found in experiments.

\begin{figure}[t!]
\centering\includegraphics[
width=85mm 
]{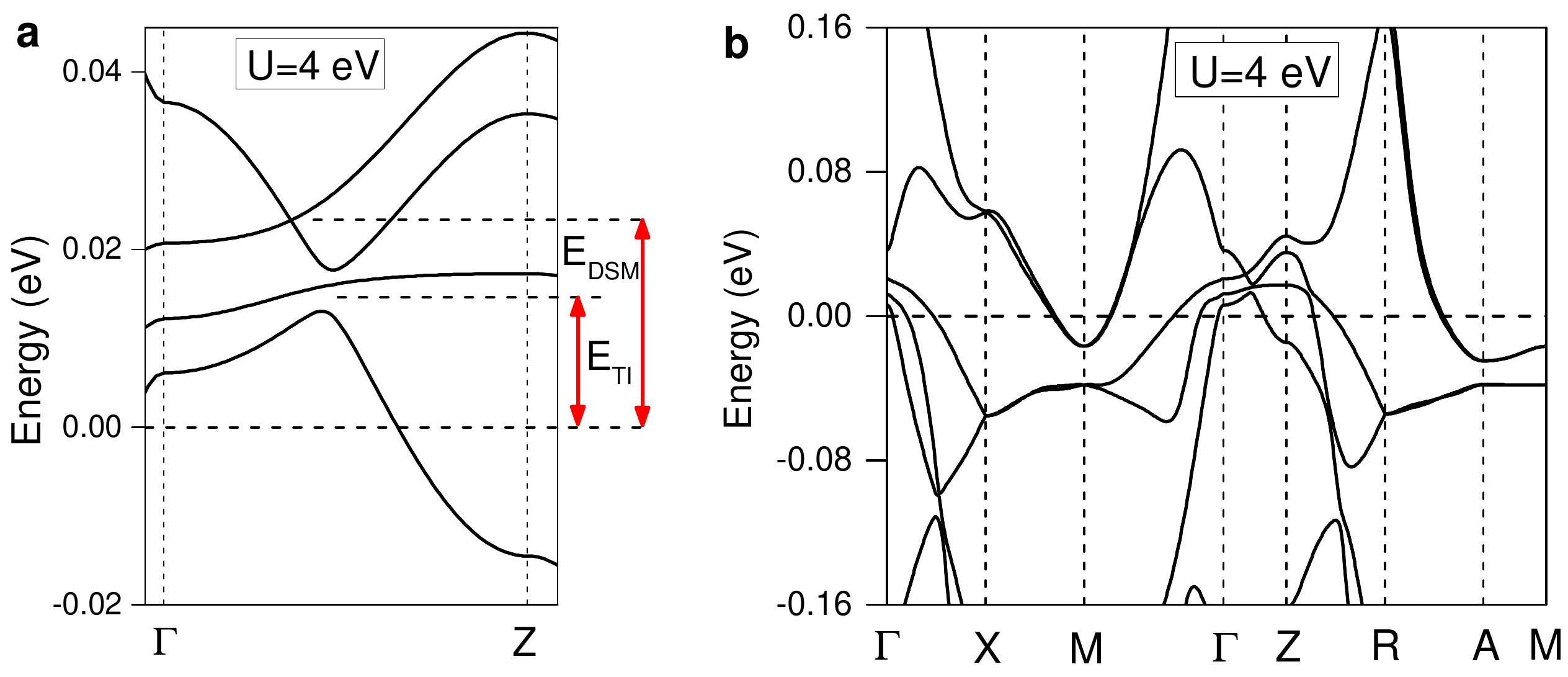}
\caption{(Color online)
{\bf a}:
Renormalized band structure near the Fermi energy $E_F$ for $U=4$ eV, along $\Gamma$-$Z$
showing the band inversion, with $E_{\rm{TI}}\approx 14$ meV and
a
Dirac
level crossing at $E_{\rm{DSM}}\approx 22$ meV.
{\bf b}: The corresponding band structure along high symmetric directions of the BZ.
}
\label{fig:3}
\end{figure}

We now consider the effects of electron correlations.
The evolution of the orbital resolved quasiparticle spectral weight $Z_{\alpha}$ with $U$,
for a typical ratio of $J_{\rm{H}}/U=0.25$, is shown in Fig.\,S2.
At $U\sim2-3$ eV, the system crosses over to a strongly correlated metal, where $Z_{\alpha}$
in all orbitals drop rapidly and
become strongly orbital selective. This feature is universal for FeSCs. In the strongly correlated metal phase,
the $d_{xy}$ orbital is the most strongly correlated, with $Z\lesssim0.1$ for $U\gtrsim4$ eV. As expected (see Fig.~\ref{fig:3}{\bf b}),
the bands are highly
 renormalized.
 We find the band inversion persists under such a strong orbital-selective correlation up to $U\approx10$ eV.
 Comparing the band structure at $U=4$ eV in Fig.~\ref{fig:3}{\bf a} to that at $U=0$ in Fig.~\ref{fig:2},
  the characteristic energies are very close to the the Fermi energy:
$E_{\rm{TI}}\approx 14$ meV and $E_{\rm{DSM}}\approx 22$ meV at $U=4$ eV.

\begin{figure}[t!]
\centering\includegraphics[
width=85mm 
]{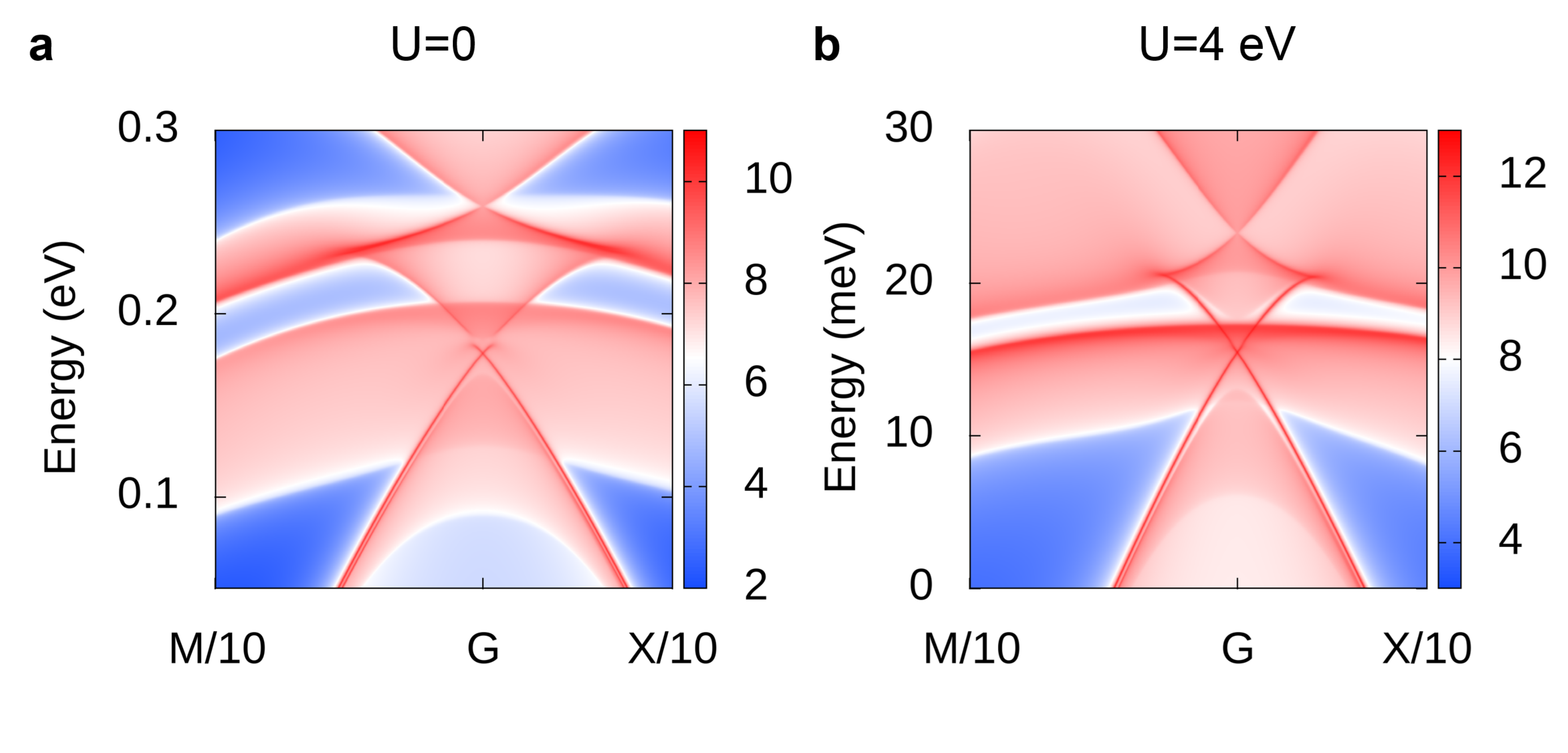}
\caption{(Color online) Topological surface states showing the Dirac cone features of the model
on the (001) surface,
at $U=0$ (in {\bf a}) and $U=4$ eV (in {\bf b}).
}
\label{fig:4}
\end{figure}

To demonstrate the band topology, we calculate the surface state spectrum
 by taking a slab-like lattice geometry. As shown in Fig.~\ref{fig:4}, for both $U=0$ and $U=4$ eV, a clear Dirac cone
 shows up in the bulk gap at the $\Gamma$ point.
 Slightly above in energy,
 there is another Dirac cone that is associated with
 the DSM feature of the bulk spectrum.
The correlation effects
 reduce the energy positions of the Dirac cones by about one order of magnitude. For $U=4$ eV,
 the calculated Dirac surface state is only about 15 meV above the Fermi level.
 A slight electron doping in the system may further push this energy to be below $E_F$.

{\it Band shifts caused by orbital-selective correlation and the robustness of band inversion.}
The orbital-selective correlations
 push the $p_z+d_{xy}^-$ band
 towards $E_F$.
This would make the energy position of this band lower than those of the $d_{xz/yz}$ bands at the $\Gamma$ point,
and hence remove the band inversion.
Nonetheless,
our calculation shows that the band inversion persists over a wide range of interaction up to about $U=10$ eV.
To understand the robustness of the band inversion in the model,
we examine the energy positions of the $p_z+d_{xy}^-$ and $d_{xz/yz}$ bands at $\Gamma$ point for $U=0$ and $U=4$ eV,
respectively. The energy of the $p_z+d_{xy}^-$ band is renormalized from $E\approx0.8$ eV to $E\approx0.04$ eV
at $U=4$ eV. The renormalization factor is about $20$, which is close to the value of $1/Z_{xy}\approx18$ at $U=4$ eV.
On the other hand, the energy of the $d_{xz/yz}$ bands (when neglecting the splitting by spin-orbit coupling)
is renormalized from $E\approx0.1$ eV to $E\approx0.01$ eV, \textit{eg}.,
by a factor of $10$. This factor is almost twice as large as $1/Z_{xz/yz}\approx6$ at $U=4$ eV.
This unusually strong renormalization of the $d_{xz/yz}$ band energy
helps keep the $d_{xz/yz}$ band lower than the $p_z+d_{xy}^-$ band at $\Gamma$ point, and hence maintain the band inversion.

Near the $\Gamma$ point, the hybridization between the $d_{xy}$ and $d_{xz/yz}$ orbitals is weak.
The energy of a band with a dominant $\alpha$ orbital
at a wave vector $k$ is then written as $E_{\alpha k} = \epsilon_{\alpha} + \xi_{\alpha}(k)$,
where $\epsilon_{\alpha}$ is the onsite energy associated with the crystal field splittings, and $\xi_{\alpha}(k)$
is the dispersive part of the band which is obtained from Fourier transformation of hopping parameters.
Under electron correlation the energy is renomalized to $\tilde{E}_{\alpha k} = Z^\prime_{\alpha} \epsilon_{\alpha} + Z_{\alpha} \xi_{\alpha}(k)$,
where $Z_{\alpha}$ is the quasiparticle spectral weight and $Z^\prime_{\alpha}$ is a renormalization factor
for the onsite energy $\epsilon_{\alpha}$~\cite{Lin_arXiv_2021}.
 In a system with strong orbital selectivity, $Z^\prime$ and $Z$ may differ
 sizeably, given that
 $Z^\prime$ reflects the redistribution of electron among different orbitals~\cite{Lin_arXiv_2021} (see Fig.~S3 of SM~\cite{SM}).
 Rewriting $\tilde{E}_{\alpha k}=Z_{\alpha} E_{\alpha k} +(Z^\prime_{\alpha}-Z_{\alpha})\epsilon_{\alpha}$,
 it
 can be seen that the difference between $Z$ and $Z^\prime$ generates an additional energy shift of a band.
 It has been shown that this leads to a reconstruction of the bandstructure and Fermi surface in LiFeAs~\cite{Lin_arXiv_2021}.
 In the case of FeSe$_x$Te$_{1-x}$, we show that this effect helps stabilize the band inversion under strong orbital selective correlation.
 By comparison, we find $Z_{xy}\approx0.056$ and $Z^\prime_{xy}\approx0.017$. Together with $\epsilon_{xy}\approx -0.137$ eV,
 this yields a small {\it positive} shift of about $0.005$ eV. For the $d_{xz/yz}$ bands, $Z_{xz/yz}\approx0.164$,
 $Z^\prime_{xz/yz}\approx-0.008$,
 and $\epsilon_{xz/yz}\approx0.051$ eV, which lead to a {\it negative} shift of about $0.009$ eV.
 This significant energy downshift accounts for the unusually strong renormalization of the $d_{xz/yz}$ band energy.
 More importantly, the energy shifts in the $d_{xy}$ and $d_{xz/yz}$ bands are in {\it opposite} directions.
 This partially cancels the effect of strong bandwidth renormalization in the $d_{xy}$ band, making the energy of the $d_{xy}$
 band at $\Gamma$ point higher than that of the $d_{xz/yz}$ bands.
  As such, the band inversion is stabilized
 over a wide range of interactions, up to about $U\approx10$ eV.

\begin{figure}[t!]
\centering\includegraphics[
width=85mm 
]{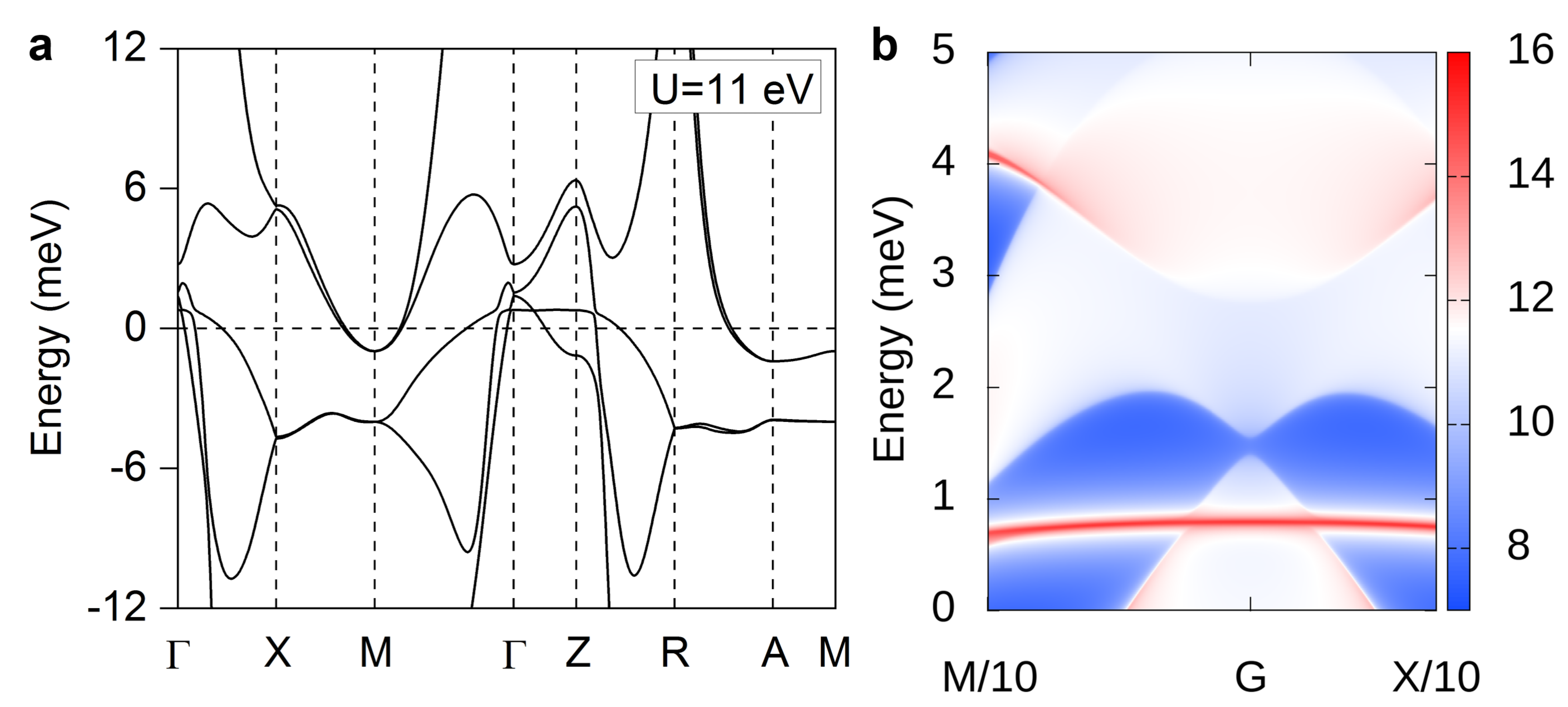}
\caption{(Color online) {\bf a}: Band structure along high symmetric directions of the BZ at $U=11$ eV, where the bands no longer invert.
{\bf b}: The corresponding surface state calculation at $U=11$ eV,
showing that the topological surface states disappear under extreme orbital selectivity.
}
\label{fig:5}
\end{figure}

{\it Topological phase transition under strong orbital selectivity.~}
Further increasing electron correlations to $U>10$ eV, the difference between $Z$ and $Z^\prime$ becomes vanishing.
In this strong orbital-selective limit, we expect the $p_z+d_{xy}^-$ band is further renormalized toward $E_F$.
Without the additional energy shifts caused by $Z^\prime$, its band energy at $\Gamma$ point should eventually
be pushed to lower than that of the $d_{xz/yz}$ bands,
and the band inversion should be removed.
We test this idea by taking $U=11$ eV, where $Z_{xy}\approx0.002$, which is very close to an OSMT.
As shown in Fig.~\ref{fig:5}{\bf a}, the $p_z+d_{xy}^-$ band is indeed lower in energy than the $d_{xz/yz}$ bands at both $\Gamma$ and $Z$ points.
As a result, the band inversion is removed and the band topology then becomes trivial.
The Dirac cone features disappear in the surface spectrum
and full gaps open at the $\Gamma$ point,
as seen in Fig.~\ref{fig:5}{\bf b}.
Such a drastic change of the band topology indicates that there is a topological phase transition under strong orbital selectivity at $U\approx11$ eV.

{\it Discussion and conclusions.~}
We have shown how orbital-selective correlations lead to a topological
band inversion and Dirac node that are
pinned to the immediate vicinity of the Fermi energy.
The strong correlations of both the $d_{xy}$ and $d_{xz/yz}$ orbitals make the band
topological features  robust over a wide range of interactions.
Our results provide a natural explanation to the observed topological surface states in FeSe$_x$Te$_{1-x}$ and other FeSCs~\cite{Zhang_NP_2019, Chen_CPL_2019}.
More generally, the cooperation of strong correlations and space-group symmetry has been advanced as a design rule for
 strongly correlated electronic topology~\cite{Chen-Natphys22}, and our work showcases
 the iron chalcogenide superconductors as a new platform for applying this principle.

\begin{acknowledgments}
We thank Lei Chen, Chandan Setty, Shouvik Sur, and Ming Yi
 for useful discussions. This work has in part been supported by
the National Science Foundation of China Grant No. 12174441.
Work at Rice was primarily supported
by
the U.S. Department of Energy, Office of Science, Basic Energy Sciences,
under Award No. DE-SC0018197, and by
the Robert A.\ Welch Foundation Grant No.\ C-1411.
Work at Los Alamos was carried out under the auspices of the U.S. DOE NNSA under Contract No. 89233218CNA000001.
It was supported by LANL LDRD Program and in part by the Center for Integrated Nanotechnologies,
a U.S. DOE BES user facility.
Q.S.acknowledges the hospitality of the Kavli Institute for Theoretical
Physics, supported in part by the National Science Foundation
under Grant No. NSF PHY-1748958, during the program
``A Quantum Universe in
a Crystal: Symmetry and Topology across the Correlation
Spectrum'',
as well as the hospitality of the Aspen Center for Physics,
which is supported by NSF grant No. PHY-1607611.

{\it Note added:}~
After completing this work, we
learnt of another work~\cite{Kim-FeTeSe23},
in which the effect of the orbital-selective Mott correlations
on the band topology of FeSe$_{0.5}$Te$_{0.5}$ is studied using the DMFT method.
\end{acknowledgments}



\clearpage
\setcounter{figure}{0}
\makeatletter
\renewcommand{\thefigure}{S\@arabic\c@figure}
\onecolumngrid

\section*{SUPPLEMENTAL MATERIAL -- Orbital-selective correlations for topology in FeSe$_{x}$Te$_{1-x}$}

\subsection{Details on the tight-binding model}


To include the realistic
band structure at low energies into our tight-binding modeling, we have first carried out band structure calculations for FeSe$_x$Te$_{1-x}$ (space group: P4/nmm) at $x=0.5$ within the framework of density functional theory (DFT).

We have used the plane wave basis set as implemented in Quantum ESPRESSO code \cite{qe_website}.
Norm-conserving pseudopotentials \cite{Hamann_PRB_2013} (PPs) and Perdew-Burke-Ernzerhof exchange-correlation functional were used in the calculations.
The disorder in FeSe$_{0.5}$Te$_{0.5}$ was described with virtual crystal approximation by mixing the PPs of Se and Te.
Experimental lattice parameters ($a=b=3.79327\;$\AA, $c=5.95519\;$\AA) \cite{Li_PRB_2009} were used in the simulations. Given that both the Fe $t_{2g}$ and Se/Te $p_z$ orbitals are relevant to the non-trivial band topology, we
fit the Wannierized bands with a 12-orbital tight-banding Hamiltonian including the 10 Fe $d$ orbitals and the 2 Se/Te $p_z$ orbitals in the Brillouin zone (BZ) corresponding to the two Fe and two Se/Te unit cell.
At this step we have used projected Wannier functions; the procedure of disentanglement was performed with the maximally-localized Wannier functions scheme as implemented in Wannier90 code \cite{Pizzi_JPCM_2020}.

\begin{figure}[h!]
\centering\includegraphics[
width=150mm 
]{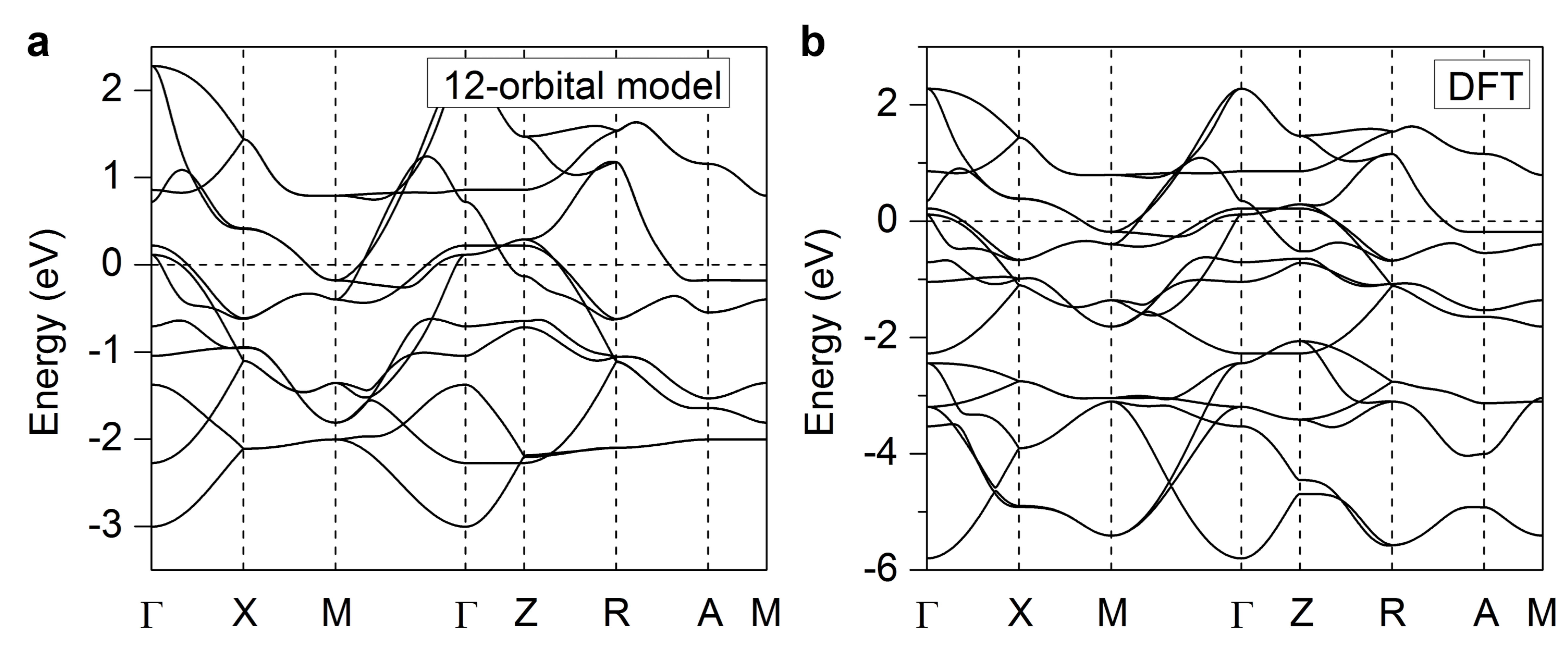}
\caption{(Color online) Band structure of FeSe$_x$Te$_{1-x}$ at $x=0.5$ along high-symmetry directions of the BZ calculated
from the 12-orbital tight-binding model (in {\bf a}) and DFT (in {\bf b}).
}
\label{fig:S1}
\end{figure}

The band structure of the 12-orbital tight-binding model compared to the DFT results is shown in Fig.~\ref{fig:S1}. The tight-binding model reproduces a similar band structure of DFT in the energy window
of interest,
from $-3$ eV to $2$ eV

\subsection{Details of the orbital-selective
correlations and the
effects on the band structure}

\begin{figure}[h!]
\centering\includegraphics[
width= 100 mm 
]{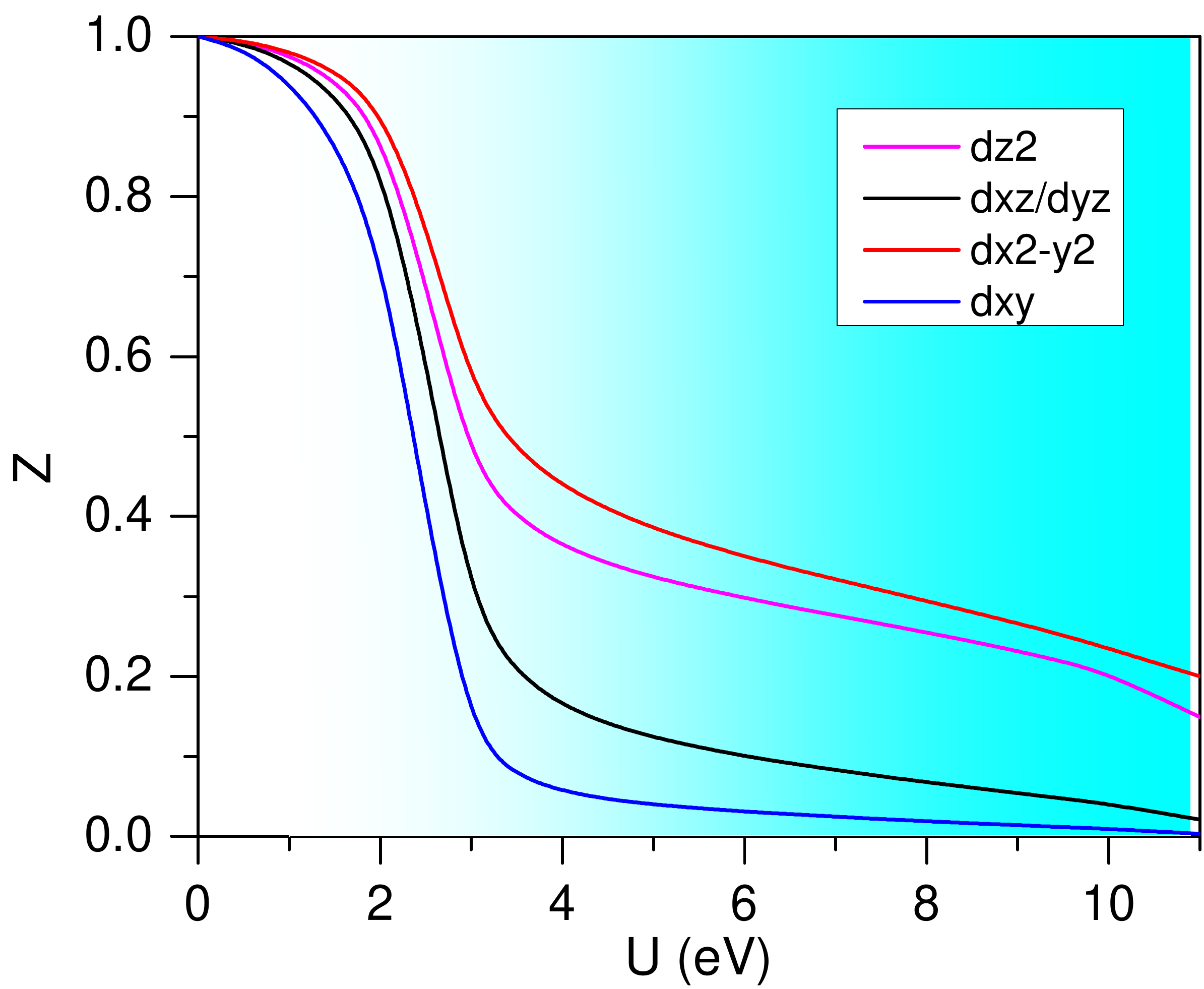}
\caption{(Color online) The orbital resolved quasiparticle spectral weight $Z$
as a function of the interaction parameter
 $U$ of the 12-orbital model. Shading shows the regime where the
 band inversion and Dirac nodes are
 pinned close to the Fermi energy.
}
\label{fig:S2}
\end{figure}

\begin{figure}[h!]
\centering\includegraphics[
width=150mm 
]{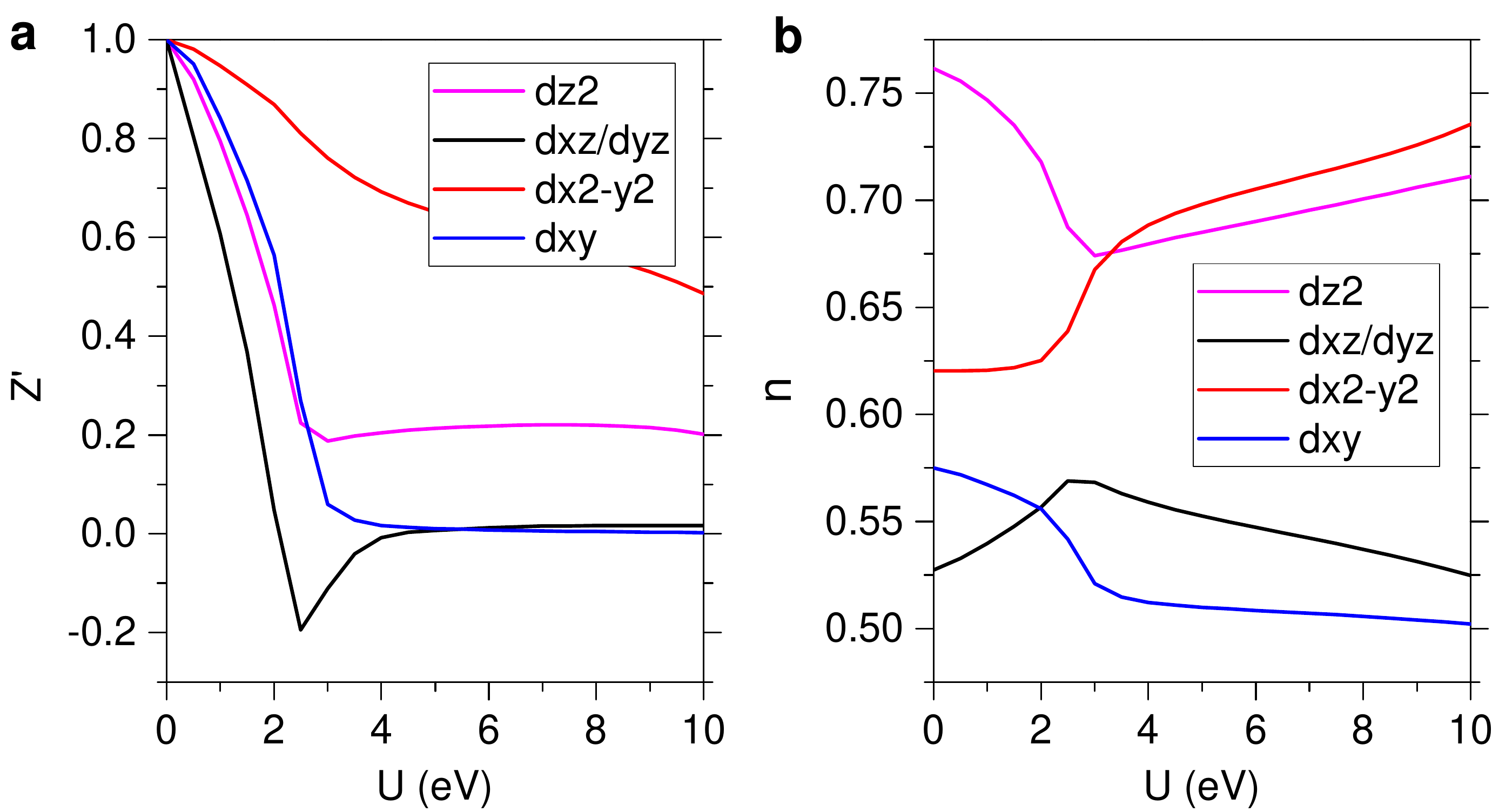}
\caption{(Color online) {\bf a}: The renormalization factor $Z^\prime$ of the onsite energy $\epsilon$
as a function of
 $U$ in the 12-orbital model.
 The renormalization is toward $E_F$, which is set to $0$. Accordingly,
 $Z^\prime$ can
 be negative,
 which corresponds to a renormalized onsite energy below $E_F$.
 {\bf b}: Electron occupation number $n$ in each orbital (per spin flavor)
 vs
  $U$ in the same model.
}
\label{fig:S3}
\end{figure}

We study the electron correlation effects of the 12-orbital Hubbard model by using the $U(1)$ slave spin theory as described in the main text. Fig.~\ref{fig:S2} shows the evolution of the quasiparticle spectral weight $Z$ of the Fe $3d$ orbitals with interaction
$U$ at $J_{\rm{H}}/U=0.25$. At $U\sim 2-3$ eV, the system undergoes a crossover from a weakly correlated metal to
a strongly correlated one identified by a rapid drop of $Z$ in all orbitals. In the strongly correlated metallic regime,
the system exhibits strong orbital selectivity, with $Z$ values in $d_{xy}$ and $d_{xz/yz}$ orbitals much smaller than those of
the $e_g$ orbitals. The crossover and strong orbital selectivity is a well known universal feature of
the multiorbital model for the iron chalcogenides, and is driven by a non-trivial interplay of
the Hund's coupling and crystal field splitting. Compared to the two $e_g$ orbitals, the three $t_{2g}$ orbitals are located at higher energies, and their occupation numbers are closer to half-filling, as shown in Fig.~\ref{fig:S3}{\bf b}.
As a result, they are closer to a Mott localized state, namely, with smaller quasiparticle spectral weights. This causes strong renormalization of both the $d_{xy}$ and $d_{xz/yz}$ bands, which is important for stabilizing the band inversion while
pinning the topological feature close to the Fermi energy. The shading in Fig.~\ref{fig:S2} shows the range of $U$ where the
 topological surface states are no more than $30$ meV away from the Fermi energy. It spans a wide range from $U\sim 3$ eV to $U\approx11$ eV in the strongly correlated metal
 regime.

As explained in the main text,
the
$d$-orbital-based bands have their widths
 renormalized by the quasiparticle spectral weight $Z$, whereas the onsite energy $\epsilon$ is renormalized by a different factor $Z^\prime$,
 which is also orbital dependent. Besides the band width renormalization, the orbital-selective correlation also leads to a redistribution of
 the electron
 density among the different orbitals, which is reflected by the renormalization of $\epsilon$.
 In general, when the electron density $n$ is increasing, the onsite energy decreases, and vice versa.
In the 12-orbital model at $U=0$, $\epsilon>0$ for the $d_{xz/yz}$ orbitals, and $\epsilon<0$ for the $d_{xy}$ orbital.
We then expect
the evolution of
$Z^\prime$
to be
in opposite to that of $n$ in the $d_{xz/yz}$ orbitals whereas $Z^\prime$ follows the behavior of $n$ in the $d_{xy}$ orbital. This is indeed the case as shown in Fig.~\ref{fig:S3}.

\begin{figure}[h!]
\centering\includegraphics[
width=150mm 
]{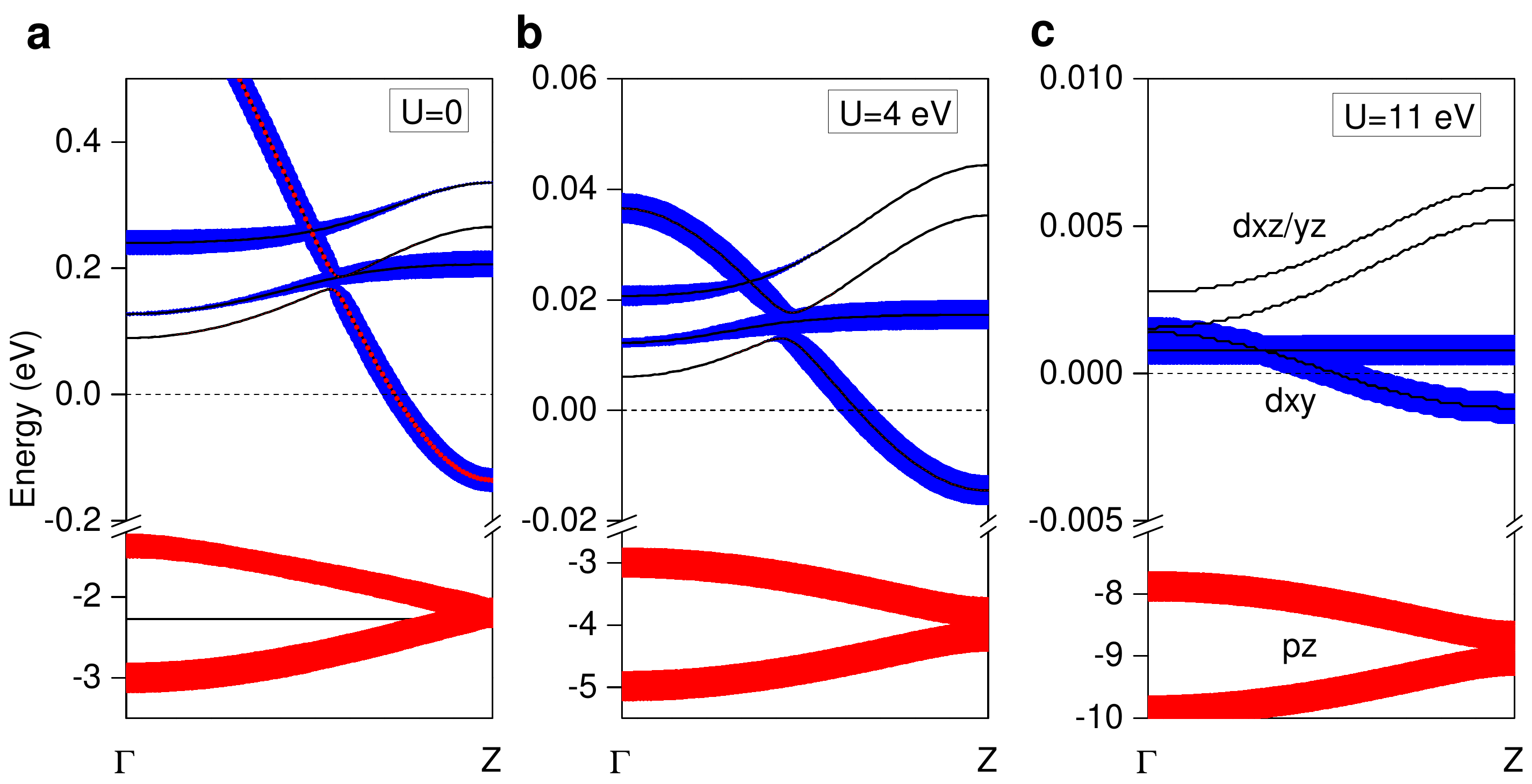}
\caption{(Color online) The band structures of the 12-orbital model for FeSe$_x$Te$_{1-x}$ at $x=0.5$ along the $\Gamma$-$Z$ direction of the BZ at $U=0$ ({\bf a}), for $U=4$ eV ({\bf b}), and for $U=11$ eV ({\bf c}), respectively. The thickness of the blue and red colors are proportional to the orbital weights of the $d_{xy}$ and $p_z$ orbitals, respectively.
}
\label{fig:S4}
\end{figure}

As already mentioned, strong orbital-selective correlations significantly affect the band structure of the 12-orbital model.
In Fig.~\ref{fig:S4}, we
 show how the bandstructure along the $\Gamma$-$Z$ direction evolves with increasing interaction $U$.
 With increasing $U$ the major effects of the electron correlation
 on the bandstructure are as follows: First, the $d$ orbital bands are strongly renormalized toward the Fermi energy $E_F$ (set to $0$ in the calculation). Second, the band inversion is a robust feature persisting up to $U\approx 10$ eV. Third, the hybridization between
 the $d$- and $p$-orbitals is
 weakened since this term is also renormalized by the factor of $\sqrt{Z}$ of the associated $d$-orbital.
 Finally, as a consequence of the decoupling between the $d$- and $p$-orbitals and renormalization of the $d$-orbital
 bands, the $p$-orbital dominant bands are shifted to lower energies with increasing $U$.

\end{document}